\documentclass[twocolumn,showpacs,preprintnumbers,amsmath,amssymb]{revtex4}
\begin{document}

\title{Faithful Teleportation with Partially Entangled States}  
\author{Gilad Gour}
\affiliation{Theoretical Physics Institute,
Department of Physics, University of Alberta,\\
Edmonton, Canada T6G 2J1}
\email{gilgour@phys.ualberta.ca}

\date{\today}

\begin{abstract} 
We write explicitly a general protocol for faithful teleportation
of a $d$-state particle (qudit) via a partially entangled pair of (pure) 
$n$-state particles. The classical communication cost (CCC) of the protocol
is $\log_{2}(nd)$ bits, and it is implemented by a {\em projective}
measurement performed by Alice, and a unitary operator performed
by Bob (after receiving from Alice the measurement result). 
We prove the optimality of our protocol by a comparison with the 
concentrate and teleport strategy. We also show that if $d>n/2$
or if there is no residual entanglement left after the faithful 
teleportation, the CCC of {\em any} protocol is at least $\log_{2}(nd)$ 
bits. Furthermore, we find a lower bound on the CCC in the process transforming 
one bipartite state to another by means of local operation and classical
communication (LOCC). 
\end{abstract}


\maketitle

In the process of quantum teleportation, one party, called Alice,
transfers an unknown quantum state to a second party's system, operated by 
Bob. There are two distinctive resources for the process: (1) The
classical information transmitted from one party to the other.
(2) The two parties share an entangled
state. In the original protocol~\cite{Bennett}, it has been shown that
the resources of $2\log_{2}d$ bits of classical information and a pair
of $d$-state particles in a maximally entangled state are sufficient
for a faithful teleportation of a $d$-state object (or qudit).   
Since then there where several generalizations for the original
protocol with more general
channels~\cite{others1,others2,others3,others4}. However, until now,   
faithful teleportation protocols (i.e. with unit fidelity and
unit probability of success~\cite{Pati,Enk}) of a $d$-state 
object have not been considered for the case when the
entangled resource is a partially entangled pair of pure $n$-state
particles (with $n>d$). In this paper, we introduce a
{\it protocol} for this scheme, and prove its optimality 
in a restricted sense by
showing that the classical communication cost (CCC) in protocols
that involves first concentration and then teleportation is at least 
$\log_{2}(nd)$ bits (which is the CCC used in our protocol).
Moreover, it is shown that if $d>n/2$ or if
there is no entanglement shared between Alice and Bob after Alice's 
measurement, the CCC of {\em any} strategy   
is at least $\log_{2}(nd)$ bits~\footnote{Here we assume that Alice
transmits her entire measurement outcome, or equivalently
that Alice and Bob do not discard any part of their state at
any time.}. We also find a lower bound on the classical
information required in the process of deterministic entanglement 
concentration~\cite{Morikoshi}.      

In the following, the qudit which is faithfully teleported from Alice
to Bob is denoted by 
\begin{equation}
|\psi _{d}\rangle _{1}=\sum_{m=1}^{d}a_{m}|m\rangle _{1}\;,
\label{alis}
\end{equation}
 and the entangled resource shared between Alice and Bob is denoted by
(not necessarily maximally entangled)
\begin{equation}
|\chi\rangle _{23}=\sum _{k=1}^{n}\sqrt{p_{k}}
|k\rangle _{2}|k\rangle _{3}\;,
\end{equation}
where $n={\rm Sch}(|\chi\rangle _{23}\rangle)$ is the Schmidt number 
(we have included in the sum only the non-zero $p_{k}$'s). Thus,
systems $1$ and $2$ belong to Alice's lab, and system $3$ to Bob's
lab. 

The teleportation can be achieved by a protocol involving just the
following steps~\cite{LP}: Alice performs a single generalized measurement on 
her systems 1 and 2, and then sends the result to Bob, who performs  
a particular unitary operation on his system 3, according to Alice's
message.   

There are two interesting questions to ask. First, what are the
conditions that the Schmidt 
numbers $\{p_{k}\}$ must satisfy in order to achieve a faithful 
teleportation (i.e. with maximum fidelity, $f=1$). Second,
what is the lower bound on the amount of classical bits 
that Alice must send to Bob.

The answer to the first question follows directly from
Nielsen's theorem~\cite{Nielsen}, and we have summarized it in the
following theorem.\\  

{\bf Theorem 1}:
Faithful teleportation is possible if, and only if, 
\begin{equation}
E_{t}(|\chi\rangle _{23})\equiv -\log _{2}p_{m} 
\geq \log_{2} d\;,
\label{t1}
\end{equation}
where $p_{m}={\rm max}\{p_{k}\}$. 
That is, teleportation is possible if, and only if, none of the
Schmidt coefficients are greater than $1/d$.
This also implies that the Schmidt number
$n$ is greater or equal to $d$.

The entanglement measure for faithful teleportation,
$E_{t}(|\chi\rangle _{23})$ (here called entanglement of
teleportation), has been defined earlier~\cite{Morikoshi} in the 
context of {\it deterministic} entanglement concentration.  
In~\cite{Morikoshi}, it has been shown (with
different notations) that $|\chi\rangle_{23}$ can be transformed 
(deterministically) by local operations and classical communications
(LOCC) to a maximally entangled pair of qudits if, and only if,
condition~(\ref{t1}) is satisfied. This provides a proof for theorem 1, 
since a maximally entangled pair of qudits can be used for a
teleportation of an unknown qudit~\cite{Bennett}.      

In order to partially answer the second question, let us first consider 
protocols that involves two steps: Alice and Bob concentrate their
entangled resource, $|\chi\rangle _{23}$, to a $d\times d$ maximally
entangled state and then teleport the state $|\psi _{d}\rangle$ (using 
the Bennett {\it et al.}~\cite{Bennett} protocol). It follows from
theorem 2 (see below) that for these protocols the CCC is at least 
$\log_{2}(nd)$ bits.

{\bf Theorem 2}:  Let $n_{1}$ and $n_{2}$ ($n_{1}\geq n_{2}$) be the
Schmidt numbers of two bipartite states $|\chi ^{(1)}\rangle _{23}$ and 
$|\chi ^{(2)}\rangle _{23}$, respectively. If $|\chi ^{(1)}\rangle
_{23}$ can be transformed to $|\chi ^{(2)}\rangle$ by LOCC, then the CCC
of the transformation is at least $\log _{2}(n_{1}/n_{2})$ bits.

{\bf Proof}: Let us write the states $|\chi ^{(1)}\rangle _{23}$
and $|\chi ^{(2)}\rangle _{23}$ in their Schmidt 
decomposition
\begin{eqnarray}
|\chi ^{(1)}\rangle _{23} & = & \sum _{k=1}^{n_{1}}\sqrt{p ^{(1)}
 _{k}}|k\rangle _{2}\otimes|k\rangle _{3}\nonumber\\
|\chi ^{(2)}\rangle _{23} & = & \sum _{m=1}^{n_{2}}\sqrt{p ^{(2)}
 _{m}}|m\rangle _{2}\otimes |m\rangle _{3}\;.
\label{chi}
\end{eqnarray}
As we have mentioned earlier, the transformation $|\chi ^{(1)}\rangle
_{23} \; \rightarrow \; |\chi ^{(2)}\rangle _{23}$ can be achieved by a
single generalized measurement performed by Alice and a unitary operation
performed by Bob (see~\cite{LP}). Let us describe Alice's measurement by the
measurement operators, $\hat{M}^{(j)}$, where $j=1,2,...,s$. That is,
\begin{equation}
\sum_{j=1}^{s}\hat{M}^{(j)\dag}\hat{M}^{(j)}=I\;,
\label{comp}
\end{equation} 
where $I$ is the identity operator. Now, after Alice obtain the outcome $j$,
the state of the system is proportional to
\begin{equation}
\hat{M}^{(j)}|\chi ^{(1)}\rangle _{23}=\sum _{k=1}^{n_{1}}\sqrt{p ^{(1)}
 _{k}}\left(\hat{M}^{(j)}|k\rangle _{2}\right)\otimes |k\rangle _{3}\;.
\end{equation}
Thus, after Bob perform a unitary operation, $\hat{u}^{(j)}$, the state
of the system, $|\chi ^{(2)}\rangle _{23}$, can be written as
\begin{equation}
|\chi ^{(2)}\rangle _{23}=(N^{j})^{-1/2} \sum _{k=1}^{n_{1}}\sqrt{p ^{(1)} _{k}}
\left(\hat{M}^{(j)}|k\rangle _{2}\right)\otimes 
\left(\hat{u}^{(j)}|k\rangle _{3}\right)\;,
\end{equation}
where $N^{j}$ is the normalization coefficient. By a comparison of the above 
equation with the expression for $|\chi ^{(2)}\rangle _{23}$ in 
Eq.~(\ref{chi}) we obtain
\begin{eqnarray}
\hat{M}^{(j)}|k\rangle _{2}=\sqrt{\frac{N^{j}}{p ^{(1)} _{k}}}
\sum _{m=1}^{n_{2}}{\;}_{3}\langle k|\hat{u}^{(j)\dag}|m\rangle _{3}
\sqrt{p ^{(2)} _{m}}|m\rangle _{2}\;.
\end{eqnarray}
That is, the operator $\hat{M}^{(j)}$ (as well as $\hat{M}^{(j)\dag}\hat{M}^{(j)}$)  
projects the $n_1$ states $|k\rangle _{2}$, into a $n_{2}$ dimensional Hilbert space.
Thus, from the completeness equation~(\ref{comp}) it follows that $sn_{2}\geq n_{1}$,
or equivalently, $\log_{2}s\geq \log _{2}(n_{1}/ n_{2})$ $\Box$. 

Note that according to theorem 2, if Alice and Bob first transform the 
state $|\chi\rangle_{23}$ to a $d\times d$ maximally entangled state
it will cost them at least $\log_{2}(n/d)$ classical bits. Adding to
it $2\log_{2}d$ bits (see Bennett {\it et al.}~\cite{Bennett}) will 
give a total of at least $\log_{2}(nd)$ classical bits for the 
concentrate and teleport strategy. The CCC of our protocol 
(see the next section) is exactly $\log_{2}(nd)$ bits. Therefore, in this 
sense our protocol is optimal. This, however, does not mean that there
are no other strategies in which the CCC is less then $\log_{2}(nd)$ bits.

For example, consider the case in which the entanglement resource
shared between Alice and Bob is given by a product of two bell states, 
i.e.
\begin{eqnarray}
&{}& |\chi\rangle _{23} = |{\rm Bell}\rangle _{23}|{\rm Bell}
\rangle_{23} \nonumber\\
&\equiv& \frac{1}{2}\left(|1,1\rangle_{23}+|2,2\rangle_{23}+|3,3\rangle_{23}
+|4,4\rangle_{23}\right)\;.
\label{ggg}
\end{eqnarray}
If Alice wishes to teleport a qubit to Bob, she
can do it with only two classical bits using one of the two Bell
states. In this case, after the teleportation, there is a residual 
entanglement left. This simple example implies that the minimum amount of 
classical information that Alice must transmits to Bob, is depending on
the residual entanglement left after the teleportation has been accomplished.

Let us denote by ${\cal E}_{r}^{(d)}(|\chi\rangle_{23})$ the {\em maximum} 
Schmidt entanglement (i.e. a logarithm of the Schmidt number) which can remain 
after a $d$-state has been faithfully teleported 
from Alice to Bob via $|\chi\rangle_{23}$. 
Note that if $|\chi\rangle_{23}$ is a 
$d$-maximally entangled state, then ${\cal E}_{r}^{(d)}(|\chi\rangle_{23})=0$. However,
there are many $n$-partially entangled states ($n\geq d$) for which 
${\cal E}_{r}^{(d)}(|\chi\rangle_{23})=0$. In particular, for $d>n/2$ the residual 
entanglement, ${\cal E}_{r}^{(d)}(|\chi\rangle_{23})$, must be zero. 

The argument goes as follows: after the teleportation, the final state of Alice and Bob 
systems can be written in the form
\begin{equation}
|{\rm final}\rangle _{123}=|{\rm RE}\rangle_{12b_{1}}|\psi_{d}\rangle_{b_{2}}\;,
\end{equation}
where $b_{1}$ is the part of Bob's system 3 that is entangled with Alice systems 1 and 2.
Therefore, the state $|{\rm RE}\rangle_{12b_{1}}$ represents the residual entanglement.
The system $b_{2}$ is the non-entangled part consisting of the teleported
state in Bob's system 3. Let us now denote the Schmidt number of 
$|{\rm RE}\rangle_{12b_{1}}$ by $n_{s}$. Since the dimension of $b_{1}$ is at least
$n_{s}$ and the dimension of $b_{2}$ is at least $d$, the dimension of Bob's
system $n\geq n_{s}d$. It is therefore clear that if $d>n/2$ then $n_{s}=1$ (i.e. zero
entanglement). Moreover,
\begin{equation}
{\cal E}_{r}^{(d)}(|\chi\rangle_{23})=-\log_{2}n_{s}\leq \log_{2}n-\log_{2}d\;.
\end{equation} 
Let us show now that if ${\cal E}_{r}^{(d)}(|\chi\rangle_{23})=0$, 
the lower bound on the amount of classical bits that Alice must send to 
Bob is given by $\log _{2}(nd)$.

Imagine teleporting a (full Schmidt number) entangled state
corresponding to the system 0-1. Alice has the system 1, and 0 is the
reference system. Alice and Bob share an entangled state, $|\chi\rangle _{23}$, 
corresponding to the system 2-3. Since Alice wants to teleport her state perfectly,
she must completely destroy the entanglement with the reference system 0.
Thus, if we assume ${\cal E}_{r}^{(d)}(|\chi\rangle_{23})=0$, she also needs to destroy all
entanglement with Bob's system 3. The dimension of the system 1-2 is $nd$,
so to disentangle it from 0-3 requires a measurement with at least $nd$
linearly independent elements, i.e. $\log _{2}(nd)$ classical bits. 

When $n=d$, our bound reduces to $2\log_{2}d$, which has been proposed 
in~\cite{Bennett} when the teleportation of a $d$-dimensional state
is performed with a $d$-maximally entangled state (i.e. with 
a Schmidt number $d$). For $n>d$ the bound is stronger
assuming there is no residual entanglement left. This means, that if
Alice and Bob have to use {\it all} of their entanglement resource in 
order to teleport the qudit, Alice will need to send at least 
$\log _{2}(nd)$ of classical bits. 
On the other hand, in the example above (see Eq.~(\ref{ggg})) $n=4$,
and therefore $\log_{2}(nd)=3$. That is, after Alice transmitted the 
two classical bits to Bob, if she wishes
also to destroy the residual entanglement
she will need to perform one more measurement (that is equivalent to 
one more classical bit).  

If $d>n/2$, ${\cal E}_{r}^{(d)}(|\chi\rangle_{23})=0$, and therefore
Alice will need to transmit Bob at least $\log_{2}(nd)$ bits of classical
information. This result is very interesting. It shows, 
for example, that if Alice and Bob share a 
$n$-maximally entangled state (with $n<2d$), Alice will have to send Bob {\it more}
classical bits then she would have to if they shared a 
$d$-maximally entangled state. This simple example emphasizes
that an increment in the entanglement of the resource will not necessarily reduce
the amount of classical bits that are indispensable for a faithful teleportation of
a qudit, but will more likely increase it.
  
Let us end this section, by showing how the lower bound of $\log_{2}(nd)$ classical bits  
leads to another bound on the minimal amount of classical communication that is required 
for the process of {\em deterministic} entanglement concentration~\cite{Morikoshi}( for
the original asymptotic entanglement concentration see~\cite{BBPS}). 
In this process, Alice and
Bob share a $d^{n}$-dimensional state $|\psi\rangle _{AB}^{\otimes n}$,
where $|\psi\rangle _{AB}$ is a partially entangled state with a
Schmidt number $d\equiv{\rm Sch}(|\psi\rangle _{AB})$. 
Suppose that by LOCC Alice and Bob transform the state into $m$-copies
of the Bell states. From~\cite{Morikoshi}, it follows that this transformation
is possible if, and only if,
\begin{equation}
m\leq nE_{t}(|\psi\rangle _{AB}). 
\label{aaa}
\end{equation}
Therefore, if this condition is satisfied, after the transformation, 
the $m$ copies of the Bell states could be used to teleport a 
$2^{m}$-dimensional state. Let us denote by $C_{1}$ the minimum amount
of classical bits that are required for the transformation 
$|\psi \rangle _{AB} ^{\otimes n}\;\rightarrow\;|{\rm Bell}\rangle
^{\otimes m}$, and by $C_{2}$ the amount that is required for the 
teleportation. Using the Bennett {\it et al.} protocol, we find that
$C_{2}=2\log_{2}2^{m}=2m$. Now, since there is no residual
entanglement left in this process, from our bound, it follows that
$C_{1}+C_{2}\geq \log _{2}\left(2^{m}d^{n}\right)$ and thus
\begin{equation}
C_{1}\geq n\log_{2}d\;-\;m\equiv nE_{Sch}(|\psi \rangle _{AB})-m\;,
\end{equation}    
where 
$E_{Sch}(|\psi \rangle _{AB})=\log_{2}{\rm Sch}(|\psi \rangle _{AB})$ 
is the Schmidt entanglement. From Eq.~(\ref{aaa}) it follows that the 
minimum bound is
\begin{equation}
C_{1}\geq n\left(E_{Sch}(|\psi \rangle _{AB})- E_{t}(|\psi\rangle _{AB}) \right).
\end{equation} 
Note that $E_{Sch}(|\psi \rangle _{AB})\geq E_{t}(|\psi\rangle _{AB})$
with equality if, and only if, $|\psi \rangle _{AB}$ is a maximally entangled state.     

\section*{A general protocol for faithful teleportation}

Let us now present a general protocol for
teleportation of a qudit with maximum fidelity ($f=1$). 
The protocol consists of a projective local measurement performed by
Alice and a subsequent unitary local operation performed by Bob.
The protocol is a general one, in the sense that Alice teleports
a qudit to Bob via $\log_{2}(nd)$ classical bits and {\em any} 
partially entangled pair of pure $n$-state particles 
that satisfy Eq.~(\ref{t1}) and Eq.~(\ref{phase}) (see below).

The protocol presented below involves $(nd)^{2}$ coefficients, $V^{(j)}_{mk}$ 
(where $j=1,2,...,nd$, $m=1,2,...,d$ and $k=1,2,...,n$), that satisfy the
following two conditions:
\begin{eqnarray}
\delta _{j'j} & = & \sum _{m=1}^{d}\sum _{k=1}^{n}V_{mk}^{(j)*}V_{mk}^{(j')}
\label{final}\\
\delta _{m,m'} & = & 
nd\sum _{k=1}^{n}p_{k}V^{(j)*}_{m'k}V^{(j)}_{mk}\;.
\label{fin}
\end{eqnarray}
As we will see later, such coefficients can be found in many cases. 
We write now the steps of
the protocol in terms of these coefficients:\\
(1) The initial state is: 
\begin{equation}
|{\rm I}\rangle_{123}\equiv|\psi _{d}\rangle _{1} |\chi\rangle _{23}\;.
\end{equation}
(2) Alice performs a joint {\em projective} measurement on systems 1 and 2; 
the corresponding projectors 
$P^{(j)}\equiv |M^{(j)}\rangle_{12\;\;12}\langle M^{(j)}|$ ($j=1,2,...,nd$)
are given in terms of the coefficients $V^{(j)}_{mk}$:
\begin{equation}
|M^{(j)}\rangle_{12}
=\sum _{m=1}^{d}\sum_{k=1}^{n}V^{(j)}_{mk}|m\rangle _{1}|k\rangle _{2}.
\label{mvv}
\end{equation}
Note that Eq.~(\ref{final}) guaranties that the $nd$ states 
$|M^{(j)}\rangle_{12}$ are orthonormal.\\ 
(3) The state of the system after Alice obtained the measurement $j$
(up to normalization):
\begin{eqnarray}
P^{(j)}|{\rm I}\rangle_{123} &=& \sum_{m=1}^{d}
\sum_{k=1}^{n}a_{m}\sqrt{p_{k}}V^{(j)*}_{mk}|M^{(j)}\rangle_{12}
|k\rangle_{3}\nonumber\\
& \equiv & \frac{1}{\sqrt{s}}\sum_{m=1}^{d}a_{m}|M^{(j)}\rangle_{12}\otimes 
\hat{u}^{(j)}|m\rangle_{3}\;,
\end{eqnarray}
where 
\begin{equation}
\hat{u}^{(j)}|m\rangle _{3}\equiv\sqrt{s}\sum_{k=1}^{n}V_{mk}^{(j)*}\sqrt{p_{k}}|k\rangle _{3}\;.
\label{dd}
\end{equation}
Eq.~(\ref{fin}) guaranties that $\hat{u}^{(j)}$ 
(as defined in the above equation) is a unitary operator; its domain of definition can
be extended to {\em all} the $n$-dimensional Hilbert space of Bob (${\cal H}^{(n)}_{3}$).\\  
(4) After Bob performs on his system 3, the unitary operation, $\hat{u}^{(j)\dag}$, 
the final (normalized) state is:
\begin{eqnarray}
|{\rm F}\rangle_{123}
&=&
\sum_{m=1}^{d}a_{m}|M^{(j)}\rangle_{12}\otimes |m\rangle_{3}\nonumber\\
&=&
|M^{(j)}\rangle_{12}\otimes|\psi_{d}\rangle _{3}\;,
\end{eqnarray}
where the teleported qudit, $|\psi _{d}\rangle _{3}$,
is given by ({\it cf} Eq.~(\ref{alis}))
\begin{equation}
|\psi _{d}\rangle _{3}=\sum_{k=1}^{d}a_{k}|k\rangle _{3}.
\label{bob}
\end{equation}
(Note that although in the above sum $k$ runs from 1 to $d$, 
${\cal H}^{(n)}_{3}$ is an $n$-dimensional Hilbert space ($n\geq d$)).
Thus, our protocol works if there are $(nd)^{2}$
parameters that satisfy both Eq.~(\ref{final}) and Eq.~(\ref{fin}).

Let us first define the $s^{2}$ parameters for the case $d=2$ and 
$n\geq 2$. This case represents a general faithful 
teleportation of a qubit. It implies that
teleportation of a qubit, if possible, can always be implemented by
a projective measurement performed by Alice and a unitary operation
performed by Bob.   
Furthermore, for the case $n=2$ we will see below that our protocol
reduces to the original one given in~\cite{Bennett}.

In the determination of the parameters $V^{(j)}_{mk}$ we will make
use of the following notations. First, 
\begin{equation}
e_{k,k'}\equiv\exp\left(i\frac{2\pi}{n}kk'\right)\;,   
\end{equation}
where $k,k'=1,2,...,n$ (note that $e_{kk'}$ is a unitary matrix).
Second, we define $n$ angles $\theta_{1},\theta_{2},...,\theta_{n}$
such that
\begin{equation}
\sum_{k=1}^{n}p_{k}\exp(i\theta _{k})\;=\;0\;.
\label{angel}
\end{equation}
Such phase factors can always be found when all the $n$ Schmidt
probabilities $p_{k}\leq 1/2$ (compare with Eq.~(24) in~\cite{Wootters}).
According to Theorem 1, for $d=2$ we have $E_{t}(|\chi\rangle _{23})\geq 1$
and therefore $p_{k}\leq 1/2$ for all $k=1,2,...,n$.

With these definitions, the protocol for $d=2$ is given by
\begin{equation}
V^{(j)}_{1k}=\frac{1}{\sqrt{s}}e_{j,k}\;\;\;
{\rm and}\;\;\;V^{(j)}_{2k}=\frac{1}{\sqrt{s}}e_{j,k}\exp(i\theta_{k})\;,
\label{11}
\end{equation}
for $1\leq j\leq n$, and  
\begin{equation}
V^{(j)}_{1k}=-\frac{1}{\sqrt{s}}e_{j,k}\exp(-i\theta _{k})\;\;\;
{\rm and}\;\;\;V^{(j)}_{2k}=\frac{1}{\sqrt{s}}e_{j,k}\;,
\label{22}
\end{equation}
for $n< j\leq 2n$.
It can be shown that these $s^{2}=4n^{2}$ parameters satisfy both
Eqs.~(\ref{final},\ref{fin}) and thus define a general protocol for
faithful teleportation of a qubit. 

Consider the case in which $n=2$, and thus $p_{1}=p_{2}=1/2$.
Two angles that satisfy Eq.~(\ref{angel}) are $\theta _{1}=0$ and
$\theta _{2}=\pi$. With this choice, Eqs.~(\ref{11},\ref{22}) yields
$V^{(2)}_{11}=V^{(3)}_{11}=V^{(j)}_{12}=V^{(2)}_{21}=V^{(4)}_{21}
=V^{(3)}_{22}=V^{(4)}_{22}=1/2$, where all the other
$V^{(j)}_{mk}=-1/2$. The 4 orthonormal measurement states are given by
(see Eq.~(\ref{mvv}))
\begin{eqnarray}
|M^{(j)}\rangle _{12} &=& V^{(j)}_{11}|1\rangle _{1}|1\rangle _{2}
+V^{(j)}_{12}|1\rangle _{1}|2\rangle _{2}\nonumber\\
&+& V^{(j)}_{21}|2\rangle _{1}|1\rangle _{2}
+V^{(j)}_{22}|2\rangle _{1}|2\rangle _{2}\;.
\end{eqnarray}  
After Bob receives the massage $j$ from Alice's measurement, he performs a
unitary operation with matrix elements
$\left(\hat{u}^{(j)\dag}\right)_{mk}=\sqrt{2}V_{mk}^{(j)}$. This
protocol is identical to the Bennett {\it et al.} one~\cite{Bennett}, if
$|\downarrow_{1}\rangle,\;|\uparrow_{1}\rangle$ in~\cite{Bennett} are identified
with $(|1\rangle _{1}\pm |2\rangle _{1})/\sqrt{2}$ and
$|\downarrow_{2}\rangle,\;|\uparrow_{2}\rangle$ are identified with 
$|1\rangle_{2},\;|2\rangle _{2}$. 
  
Let us now consider another example, in which the state shared between
Alice and Bob is given by
\begin{equation}
|\chi\rangle _{23}=\sqrt{1\over 2}|1\rangle _{1}|1\rangle _{2}
+\sqrt{1\over 3}|2\rangle _{1}|2\rangle _{2}
+\sqrt{1\over 6}|3\rangle _{1}|3\rangle _{2}
\end{equation}
According to Theorem 1, this state can be used for a
teleportation of a qubit. According to our protocol, there are 6
possible outcomes in the projective measurement performed by
Alice. Three angles that satisfy Eq.~(\ref{angel}) are $\theta _{1}=0$
and $\theta _{2}=\theta_{3}=\pi$. Substituting these values for
$\theta _{k}$ in Eqs.~(\ref{11},\ref{22}) gives 
\begin{eqnarray}
&{}&V^{(j)}_{11}=V^{(j)}_{21}={1\over\sqrt{6}}\exp
\left({2\pi j\over 3}\right)\nonumber\\
&{}& V^{(j)}_{12}=-V^{(j)}_{22}={1\over\sqrt{6}}\exp
\left({4\pi j\over 3}\right)\nonumber\\
&{}& V^{(j)}_{13}=-V^{(j)}_{23}={1\over\sqrt{6}}\;,
\end{eqnarray} 
for $j=1,2,3$, and for $j=4,5,6$,
\begin{eqnarray}
&{}& V^{(j)}_{11}=-V^{(j)}_{21}=-{1\over\sqrt{6}}\exp
\left({2\pi j\over 3}\right)\nonumber\\
&{}& V^{(j)}_{12}=V^{(j)}_{22}={1\over\sqrt{6}}\exp
\left({4\pi j\over 3}\right)\nonumber\\
&{}& V^{(j)}_{13}=V^{(j)}_{23}={1\over\sqrt{6}}\;.
\end{eqnarray}
Thus, substitution of the above values in Eq.~(\ref{mvv}) and
Eq.~(\ref{dd}) yields the 6 orthonormal states, 
$|M^{(j)}\rangle _{12}$, and the 6 unitary operators, $\hat{u}^{(j)}$.  
This determines the protocol explicitly.  
We now present the more general scheme with general
$d\geq 2$ and $n\geq d$.  

We first define $s=nd$ angles, $\theta_{mk}$,
such that
\begin{equation}
\sum_{k=1}^{n}p_{k}\exp\left[i(\theta _{mk}-\theta
  _{m'k})\right]\;=\;\delta _{mm'}
\label{phase}
\end{equation}
(recently, these factors have been used in the construction of general deterministic 
protocols for dense coding~\cite{Shay}). 
It can be shown that if such phase factors can be found, then 
$p_{k}\leq 1/d$ for all $k=1,2,...,n$. For $d=2$ and $d=n$ such phase
factors can always be found as long as $p_{k}\leq 1/d$. 
For $2<d<n$, in general, it is not {\em always} possible to find such phase 
factors~\cite{Sh}, 
but there are several cases in which one can calculate them 
explicitly~\footnote{For example, consider the case in which the set of 
$n$ probabilities $\{p_{k}\}$ can be divided into $d$ subgroups such that 
the sum of the probabilities in each subgroup is $1/d$. Then, $\theta_{mk}=\frac{2\pi}{d}ml$
if $k$ belong to the subgroup $l$ ($l=1,2,...,d$).}.
Now, according to Theorem 1,  
$E_{t}(|\chi\rangle _{23})\geq \log_{2}d$,
and therefore, $p_{k}\leq 1/d$ for all $k=1,2,...,n$. 

With these
notations (and with the assumption that the phase factors
in Eq.~(\ref{phase}) can be found) the protocol is given by 
\begin{equation}
V^{(j)}_{mk}=\frac{1}{\sqrt{s}}\exp(i\theta _{mk})
\exp\left[ij\left(\frac{2\pi}{s}m+\frac{2\pi}{n}k\right)\right]\;.
\end{equation}
It can be shown that these $s^{2}=(nd)^{2}$ parameters satisfy both
Eqs.~(\ref{final},\ref{fin}) and thus define a protocol for
faithful teleportation of a qudit. 

In conclusion, we have found lower bounds on the amount of classical
information that are required for general faithful teleportation
schemes and a deterministic entanglement concentration. We have also
found a specific protocol for faithful teleportation of a qudit, which
generalizes the protocol given in~\cite{Bennett} for the case in which
Alice and Bob share a partially entangled resource. 
The protocol requires no more
classical communication than is conceivable with a 'concentrate and
teleport' strategy. 
The next step in this direction would be to find a protocol for
teleportation using a {\em mixed} state entangled resource.  

\begin{acknowledgments}
I would like to extend my sincere gratitude to Sam Braunstein, for reviewing
this work in its preliminary stages, and for his  excellent input regarding
areas of improvement. I would also like to thank Aram Harrow for useful 
comments and help. The author is grateful to the Killam Trust for its 
financial support. 
\end{acknowledgments}


\begin{thebibliography}{100}
\bibitem{Bennett}
C.~H.~Bennett, G.~Brassard, C.~Crepeau, R.~Jozsa, A.~Peres and 
W.~K.~Wootters, Phys. Rev. Lett. {\bf 70}, 1895 (1993).

\bibitem{others1}
L.~Vaidman, Phys. Rev. A {\bf 49}, 1473 (1994).

\bibitem{others2}
S.~L.~Braunstein and H.~J.~Kimble, Phys. Rev. Lett. {\bf 80}, 869 (1998).

\bibitem{others3}
M.~Horodecki, P.~Horodecki and R.~Horodecki, Phys. Rev. A {\bf 60},
1888 (1999).

\bibitem{others4}
I. Devetak, A.W. Harrow and A. Winter, arXive e-print quant-ph/0308044.

\bibitem{Pati}
P.~Agrawal and A.~K.~Pati Phys. Lett. A {\bf 305}, 12 (2002).

\bibitem{Enk}
S.~J.~van Enk, Phys. Rev. Lett. {\bf 91}, 017902 (2003).

\bibitem{Morikoshi}
F.~Morikoshi and M.~Koashi, Phys. Rev. A {\bf 64}, 022316 (2001). 

\bibitem{LP}
H.-K. Lo and S.~Popescu, arXive e-print quant-ph/9707038.

\bibitem{Nielsen}
M. A. Nielsen, Phys. Rev. Lett. {\bf 83}, 436 (1999).

\bibitem{BBPS}
C.~H.~Bennett, H.~J.~Bernstein, S.~Popescu and B.~Schumacher, 
Phys. Rev A {\bf 53}, 2046 (1996).

\bibitem{Wootters}
W.~K.~Wootters, Phys. Rev. Lett. {\bf 80}, 2245 (1998).

\bibitem{Shay}
S. Mozes, B. Reznik and J. Oppenheim, quant-ph/0403189.

\bibitem{Sh}
S. Mozes, private communication

\end{thebibliography}
\end{document}